\documentstyle[preprint,aps,epsf]{revtex}
\newif\iftightenlines\tightenlinesfalse
\tightenlines\tightenlinestrue

\def\te{\tilde e}

\def\tq{\tilde q}
\def\tu{\tilde u}

\def\td{\tilde d}

\def\ttau{\tilde \tau}

\def\tell{\tilde\ell}
\def\tq{\tilde q}
\def\tw{\widetilde W}

\def\dofig#1#2{\epsfxsize=#1\centerline{\epsfbox{#2}}}
\begin{document}
\draft
\preprint{\vbox{\baselineskip=14pt%
   \rightline{FSU-HEP-001206}
   \rightline{UH-511-977-00}
}}
\title{Probing Slepton Mass Non-Universality at $e^+e^-$ Linear Colliders
}
\author{Howard Baer$^1$, Csaba Bal\'azs$^2$, Stefan Hesselbach$^2$, J. Kenichi
Mizukoshi$^2$ and Xerxes Tata$^{2}$}
\address{
$^1$Department of Physics,
Florida State University,
Tallahassee, FL 32306 USA
}
\address{
$^2$Department of Physics and Astronomy,
University of Hawaii,
Honolulu, HI 96822 USA
}
%
\maketitle
\begin{abstract}
There are many models with non-universal soft SUSY breaking sfermion
mass parameters at the grand unification scale. Even in the mSUGRA
model scalar mass unification might occur at a scale closer to $M_{Planck}$, 
and renormalization effects would cause a mass splitting at $M_{GUT}$.  We
identify an experimentally measurable quantity $\Delta$ that correlates
strongly
with $\delta m^2 \equiv m^2_{\te_R}(M_{GUT})-m^2_{\te_L}(M_{GUT})$, and
which can be measured at electron-positron colliders provided both
selectrons and the chargino are kinematically accessible. We show that
if these sparticle masses can be measured with a precision of 1\% at a
500~GeV linear collider, the resulting precision in the determination of
$\Delta$ may allow
experiments to distinguish between scalar mass unification at the GUT
scale from the corresponding unification at $Q\sim M_{Planck}$. Experimental
determination of $\Delta$ would also provide a distinction
between the mSUGRA model and the recently proposed gaugino-mediation
model. Moreover, a measurement of $\Delta$ (or a related quantity
$\Delta'$) would allow for a direct determination of $\delta m^2$.

\end{abstract}

\medskip


Theories with multiple scalar fields such as weak scale supersymmetry
(SUSY) contain new sources of flavour violation, and so are strongly
constrained by experiment. There are, however, a number of mechanisms to
suppress these unwanted effects. The most obvious way is to assume that
the scalars are heavy. However, heavy scalars potentially destabilize
the electroweak symmetry breaking (EWSB) sector, and are generally an
anathema within the SUSY context\footnote{We do not mean to imply that
models with heavy scalars are not allowed. Indeed a construction of
phenomenologically viable models with heavy scalars and an examination
of their viability has received considerable attention in recent years
\cite{heavy}.}. Flavour violation from supersymmetric gauge (and
gaugino) interactions is also suppressed if fermion and sfermion mass
matrices are aligned, or scalars (at least those with the same gauge
quantum numbers) are sufficiently degenerate. Indeed within the
extensively studied minimal supergravity (mSUGRA) and gauge-mediated
SUSY breaking (GMSB) models, the latter mechanism is used to suppress
these flavour violating effects. This is also the case in the more
recently proposed anomaly-mediated SUSY breaking (AMSB) and
gaugino-mediated SUSY breaking models.\footnote{For a general overview
of models with non-universal SUSY breaking terms, see Ref. \cite{jhep}.}

The mSUGRA Grand Unified Theory (GUT) framework is characterized by the
assumption of universality of soft SUSY breaking scalar masses ($m_0$)
and trilinear scalar couplings ($A_0$), renormalized at some high scale
in between $M_{GUT}$ and $M_{Planck}$. In practice, this scale is
frequently taken to be $M_{GUT}$, though it should be kept in mind that
renormalization group evolution from the true scale of unification may
lead to significant splitting of scalar masses at $Q=M_{GUT}$
\cite{pp}. A similar situation obtains in the simplest gaugino-mediation
model \cite{gm} based on $SU(5)$ where it is assumed that soft SUSY
breaking scalar masses essentially vanish at the so-called
compactification scale which is taken to be between $M_{GUT}$ and
$M_{Planck}$. The situation in these scenarios is in sharp contrast to
the corresponding situation in GMSB or AMSB models, where there may be a
diversity of soft SUSY breaking scalar masses at the high scale.

If the two selectrons and the lighter chargino $\tw_1$ are discovered
and their masses along with the mass of the lightest supersymmetric
particle (LSP) determined, it would be easy to differentiate mSUGRA,
GMSB and AMSB models from one another.  Within the GMSB framework,
$\te_L$ is considerably heavier than $\te_R$ while in the AMSB
framework, the lighter chargino is essentially degenerate with the
$SU(2)$ gaugino-like neutralino LSP.  However, differentiating between
the mSUGRA model\footnote{In the rest of the paper we will define the
mSUGRA framework to have universal parameters at $Q=M_{GUT}$.} with $m_0
\sim {1 \over 2} m_{1/2}$ and the $SU(5)$ model with gaugino mediation
of SUSY breaking is obviously more difficult. Although the conceptual
foundations of the two models are very different, they differ
quantitatively only due to the additional small splitting between the
two selectron masses at the GUT scale. Of course, such a splitting could
also arise if the scale of unification of scalar masses differs from
$M_{GUT}$; for the same value of $m_{1/2}$, such a model cannot be
readily differentiated from the gaugino mediation scenario. It is also
possible that even within the SUGRA framework $m(\te_L)$ and $m(\te_R)$
are split by $D$-term contributions from additional $U(1)$ factors of
the underlying gauge group \cite{dterm}. The purpose of this paper is to
examine how well we can probe deviations from universality of
intra-generational slepton mass parameters expected within the mSUGRA
framework, if sleptons and charginos are discovered at an
electron-positron collider operating at $\sqrt{s}=500$~GeV and their
masses are measured with a precision of $\sim 1\%$
\cite{okada,munroe,snow,zdr}.

Neglecting electron Yukawa couplings, the one-loop renormalization group
equation (RGE) for the difference of selectron masses can be written as,
\begin{eqnarray}
\frac{d}{dt}\left(m_{\te_R}^2-m_{\te_L}^2\right) = 
-{2\over {16\pi^2}}\left({9\over 5}g_1^2M_1^2-3g_2^2M_2^2 -{9
\over{10}}g_1^2S \right) \nonumber \\
=
-{1\over {2\pi}}{{M_2^2}\over \alpha_2^2}\left({9\over
5}\alpha_1^3-3\alpha_2^3\right) + {9\over {20\pi}}\alpha_1 S & &,
\label{rge}
\end{eqnarray}
where $t =\ln(Q/M_{GUT})$, and 
\begin{displaymath}
S=m_{H_u}^2-m_{H_d}^2 +
\sum(m_{\tq_L}^2-m_{\tell_L}^2 -2m_{\tu_R}^2+m_{\td_R}^2+m_{\tell_R}^2).
\end{displaymath}
Here, the sum extends over all the generations. In the last step of
Eq.~(\ref{rge}), we have assumed that gauge couplings and gaugino masses
unify at $M_{GUT}$.  The running gauge coupling $\alpha_i(Q)$ is given
at the 1-loop level by,
\begin{equation}
\alpha_i(Q) = \frac{\alpha_i(M_Z)}{1-{{b_i}\over {2\pi}}\alpha_i(M_Z)\ln({Q
\over {M_Z}})},
\label{alpha}
\end{equation}
with $b_i$ denoting the
coefficient of the 1-loop $\beta$-function for the
$i^{th}$ factor of the gauge group: $b_1=33/5$ and $b_2=1$.  
The quantity $S$ that appears above satisfies the 1-loop RGE,
\begin{equation}
\frac{dS}{dt}={{b_1}\over {2\pi}}\alpha_1 S,
\label{Srge}
\end{equation}
which can be easily integrated to give,
\begin{equation}
S(Q) = \frac{S_{GUT}}{1-{{b_1}\over {2\pi}}\alpha_1(M_{GUT})\ln({Q
\over {M_{GUT}}})} = S_{GUT}{{\alpha_1(Q)} \over \alpha_1(M_{GUT})}.
\label{Ssoln}
\end{equation}
Notice that at the 1-loop level $S_{GUT}$, and hence $S$, vanishes
within the mSUGRA framework.
It is now straightforward to integrate Eq.(\ref{rge})
between $Q=M_{GUT}$ and $Q=m_{\te}$. Noting that the factor
$M_2^2/\alpha_2^2$ on the right hand side is independent of $t$ (at one
loop), we only need to evaluate,
\begin{displaymath}
\int dt \, \alpha_i^3(t) = {{\pi} \over
{b_i}}\left(\alpha_i^2(m_{\te})-\alpha_i^2(M_{GUT})\right), 
\end{displaymath}
 to integrate the first term. We obtain,
\begin{eqnarray}
m_{\te_R}^2-m_{\te_L}^2+{{M_2^2}\over {2\alpha_2^2(M_2)}} \left[{3\over
11}\left(\alpha_1^2(m_{\te})-\alpha_1^2(M_{GUT})\right) -3\left(\alpha_2^2(m_{\te})
-\alpha_2^2(M_{GUT})\right)\right] = \nonumber \\ \delta m^2 + ({1\over
2}-2\sin^2\theta_W)M_Z^2\cos2\beta -{9\over{10b_1}}S_{GUT}\left(1 -
{{\alpha_1(m_{\te})} \over \alpha_1(M_{GUT})}\right), & &
\label{main}
\end{eqnarray}
where $\delta m^2 \equiv m_{\te_R}^2(M_{GUT})-m_{\te_L}^2(M_{GUT})$ as
well as $S_{GUT}$ vanish
within the mSUGRA framework.  The slepton masses on the left hand side
of Eq.~(\ref{main}) are the running masses evaluated at the slepton mass
scale, and are to a very good approximation the physical masses of the
selectrons.  The middle term on the right hand side of (\ref{main}) comes
from the usual hypercharge $D$-terms.

We note that except for $M_2$, all the quantities on the left-hand side
of Eq.~(\ref{main}) can be directly measured, or like the running gauge
couplings, directly obtained from measured experimental
quantities\footnote{The scale $M_{GUT}$ is defined to be the scale at
which the couplings $\alpha_1$ and $\alpha_2$ meet.}  using
Eq.~(\ref{alpha}). Furthermore, the right hand side of Eq.~(\ref{main})
is very small ($\sim 0.04 M_Z^2\cos2\beta \alt (20 \ {\rm GeV})^2$) if
scalar masses are universal, but not necessarily so otherwise.

Within the mSUGRA framework, the lighter
chargino is essentially always dominantly an SU(2) gaugino. This
suggested to us that we should examine the ``directly measurable quantity'',
\begin{equation}
\Delta =  
m_{\te_R}^2-m_{\te_L}^2+{{m_{\tw_1}^2}\over {2\alpha_2^2(m_{\tw_1})}}
\left[{3\over 11}\left(\alpha_1^2(m_{\te})-\alpha_1^2(M_{GUT})\right)
-3\left(\alpha_2^2(m_{\te}) -\alpha_2^2(M_{GUT})\right)\right],
\label{delta}
\end{equation}
which should also be small within mSUGRA as long as, ({\it i})~the
chargino mass is not very different from $M_2$, and ({\it ii})~higher
loop contributions to the RGEs are not large. It should, of course, be
kept in mind that even if the difference in ({\it i}), or the change in
sparticle masses in ({\it ii}), is at the level of just 1-2\%, the near perfect
cancellations implied by Eq.~(\ref{main}) will no longer obtain,
resulting in a large {\it relative} change in $\Delta$ from its 1-loop
value of $\sim 0.04 M_Z^2\cos2\beta \alt (20 \ {\rm
GeV})^2$. Nevertheless, even with these corrections, we expect $|\Delta|
\ll m_{\te_L}^2, m_{\te_R}^2, m_{\tw_1}^2$. Just how well $\Delta$ can
be used to measure the (non-)universality of slepton masses forms the
remainder of this paper.

\begin{figure}
\dofig{6in}{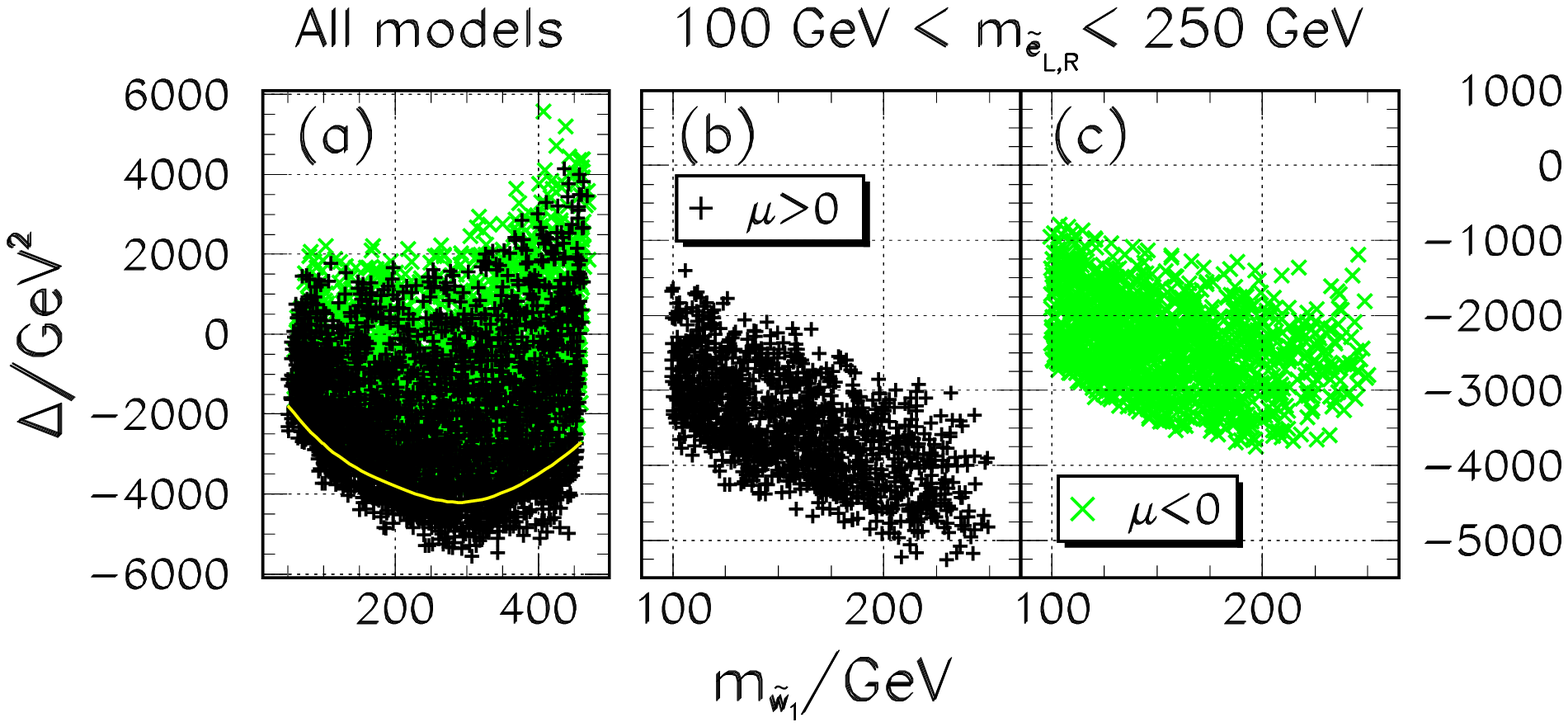}
\caption[]{A scatter plot of $\Delta$, defined in Eq.~(\ref{delta}), versus
the chargino mass. In frame ({\it a}) we show $\Delta$ for all the
generated mSUGRA models that
satisfy the experimental and theoretical constraints discussed in the
text. The light crosses and dark pluses denote models with negative and
positive values of $\mu$, respectively. Where these overlap, just the
dark pluses are visible. The white curve shows the boundary of the
band with $\mu < 0$. Frame ({\it b}) shows the same thing except that
$\te_L$, $\te_R$ and $\tw_1$ are each also required to be lighter than
250~GeV, with $\mu > 0$. Frame ({\it c}) is the same as frame ({\it b})
except that $\mu <0$. Notice that the vertical scale is on the right for
frames ({\it b}) and ({\it c}).}
\label{sugra}
\end{figure}

We begin by first examining the range of $\Delta$ within the mSUGRA
framework. Toward this end, we randomly generate mSUGRA models within
the parameter range,
\begin{eqnarray}
0 &< m_0 &< 600~{\rm GeV}, \nonumber \\
100~{\rm GeV} &< m_{1/2} &< 600~{\rm GeV}, \nonumber \\
2 & < \tan\beta &< 50, \label{range} \\
-2m_0 &< A_0 &< 2m_0, \nonumber \\
\mu &= +, - &, \nonumber
\end{eqnarray}
and compute the sparticle masses using the two loop RGEs incorporated
into the program ISAJET \cite{isajet}. We first check whether these
satisfy the lower limits on sparticle masses obtained from experiments
at LEP.  We take these to be 100~GeV for selectrons or charginos, 85~GeV
for $\ttau_1$ and 90~GeV for $h$ the lighter $CP$ even scalar
\cite{igo}.  Moreover, if $\tan\beta \leq 8$, we require $m_h \ge
112$~GeV \cite{igo}. We only keep models that ({\it i})~satisfy these
experimental constraints, ({\it ii})~break electroweak symmetry
radiatively, and ({\it iii})~have a neutralino LSP in our analysis. For
each model that we retain, we compute $\Delta$ from the chargino and
selectron masses and the {\it 1-loop couplings} as given by
Eq.~(\ref{alpha}), and show this versus the chargino mass in the
scatter plot in Fig.~\ref{sugra}{\it a}. We show a model with $\mu > 0$
($\mu < 0$) by a dark plus (light cross). Where the light crosses and
dark pluses overlap, just the latter are visible. The ``parabolic
curve'', however, shows the lower limit of $\Delta$ for models with
negative $\mu$. In Fig.~\ref{sugra}{\it b}, we once again show $\Delta$
versus the chargino mass, but this time restrict ourselves to models
with $\mu > 0$ where both selectrons and the lighter chargino are
lighter than 250~GeV, {\it i.e.} kinematically accessible at a 500~GeV
$e^+e^-$ collider that is being considered for construction some time in
the future. Fig.~\ref{sugra}{\it c} shows the same scatter plot, but for
negative values of $\mu$.

We see that the magnitude of $\Delta$ is small compared to the scale of
particle masses as we had anticipated.  Furthermore, restricting our
attention to frames {\it b} and {\it c}, {\it i.e.} to the situation
where $\Delta$ would be measurable at a 500~GeV $e^+e^-$ collider, we
see that the range of $\Delta$ is considerably diminished. Indeed models
with positive $\Delta$ appear to be possible only for heavy sleptons,
which may only be accessible at a higher energy collider. Much more
interesting to us is the fact that in frames {\it b} and {\it c}, the overlap
(for a given chargino mass) in the allowed range of $\Delta$ between
models with positive and negative $\mu$ is much reduced. We will use
this feature later in our analysis.

To get a better feel for the quantity $\Delta$, we have examined the
various factors that cause it to differ from zero, its expected 1-loop
value (aside from the small $D$-term) within the mSUGRA framework.  For
frames {\it b} and {\it c}, by far the largest factor is the replacement
of $M_2$ in Eq.~(\ref{main}) by the chargino mass.  The use of two loop
RGEs to evaluate sparticle masses yields a spread in $\Delta$ that is,
depending on the chargino mass, 5-10 times smaller than the spread in
Fig.~\ref{sugra}{\it b} or {\it c}. Moreover, since the limit on $m_h$
essentially excludes $\tan\beta \leq 4$, $\cos2\beta \alt -0.9$, and
hence, the $D$-term contribution lowers the value of $\Delta$ by an
essentially constant value of about 300~GeV$^2$, regardless of other
parameters. This then implies that if $M_2$ can be reliably determined
from the data, the expected spread in $\Delta$ will be much
reduced\footnote{This is only true for models with sleptons lighter than
250~GeV. For the case of all models shown in Fig.~\ref{sugra}{\it a},
the spreads due to two loop contributions (with scalars heavy) is
comparable to that from the approximation $M_2\simeq m_{\tw_1}$.} from
that shown in the figure.

We now turn to an examination of how well $\Delta$ can be used to
discriminate mSUGRA from models with non-universal slepton masses at
$M_{GUT}$. To introduce such a non-universality, we modify the mSUGRA
boundary conditions by choosing the GUT scale mass parameter for all
three flavours of right handed sleptons to be $m_{\tell_R}^2 =
m_0^2+\delta m^2$, but leave all other scalar masses at $m_0$. This then
ensures that no dangerous lepton flavour violation is induced
\cite{bh}. For the simple parametrization that we have introduced above,
$S_{GUT}$ in Eq.~(\ref{main}) is just $3\delta m^2$. For $\delta m^2=0$,
our model reduces to the mSUGRA model with unification of scalar masses
at $Q=M_{GUT}$.

The new parameter $\delta m^2$ that we have introduced is positive for
an mSUGRA $SU(5)$ model where scalar masses unify at a scale $Q >
M_{GUT}$. It is also positive for the gaugino-mediated SUSY breaking
model. However, since we do not want to commit to any specific form for
the underlying physics at the high scale, we analyze both signs of
$\delta m^2$.

\begin{figure}
\dofig{6in}{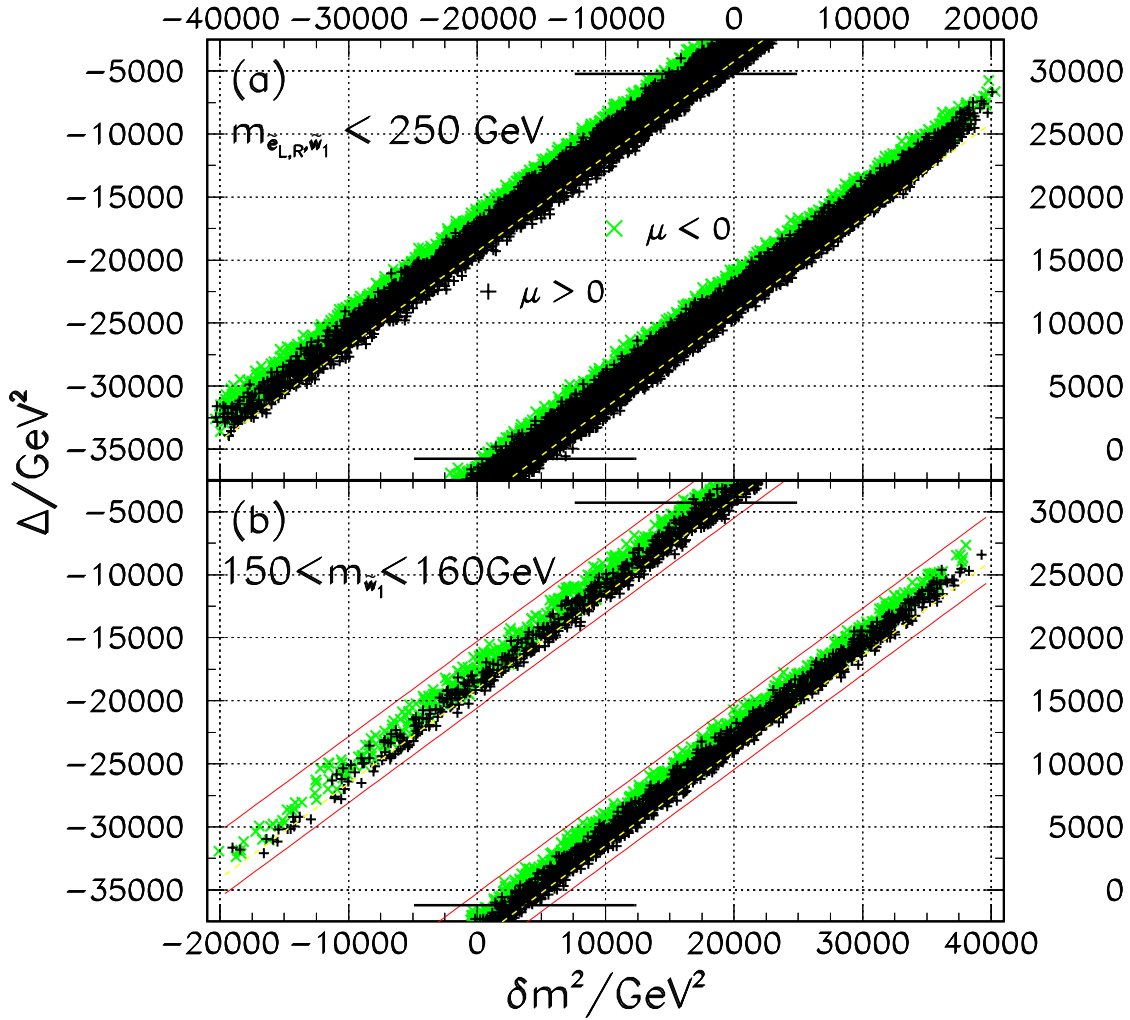}
\caption[]{A scatter plot of $\Delta$ versus $\delta m^2 \equiv
m^2_{\te_R}(M_{GUT})- m^2_{\te_L}(M_{GUT})$. In frame ({\it a}) we show
$\Delta$ for the non-universal models satisfying the same experimental
and theoretical constraints as in Fig.~\ref{sugra}{\it b,c}. Models with
positive (negative) values of $\mu$ are shown by a dark plus (light
cross). Where these overlap, just the dark pluses are visible. The
dashed line shows the lower boundary of the region with light
crosses. To expand the width of the band in which all the models lie, we
have broken the scale into two. The upper band shows $\Delta$ for
negative values of $\delta m^2$ shown on the upper scale, while the
lower band shows $\Delta$ (vertical scale on the right) for positive
values of $\delta m^2$ (lower scale). The horizontal lines show the
limits on $\Delta$ within the mSUGRA framework. Frame ({\it b}) shows
the same scatter plot as in frame ({\it a}), except that the chargino
mass is also required to lie between 150~GeV and 160~GeV. The solid
lines show the boundaries of the band in frame ({\it a}) above while the
dashed line shows the boundary of the new region with $\mu < 0$.  }
\label{nonuni}
\end{figure}

To study the effect of slepton mass non-universality on $\Delta$, we have
randomly generated such models, taking $-(200 \ {\rm GeV})^2 \leq \delta m^2
\leq (200 \ {\rm GeV})^2$, with other parameters in the range given by
(\ref{range}). 
We require that at $Q = M_{GUT}$, ({\it i})~$m_{\tell_R}^2 > 0$ 
and ({\it ii})~$m_{\tell_R}-m_{\tell_L} \leq 100$~GeV. As before, we only
accept models that satisfy the experimental and theoretical constraints
discussed previously, and for which the two selectrons and the lighter
chargino are all lighter than 250~GeV. We then compute $\Delta$ for each
of the models that we accept and plot the result versus the
non-universality parameter $\delta m^2$ in Fig.~\ref{nonuni}{\it a}. As
in Fig.~\ref{sugra}, we show models with positive (negative) values of
$\mu$ by dark pluses (light crosses). 
The crosses are again not visible when they lie under the pluses: the dashed
line marks the boundary below which there are no crosses. To 
show the spread of the models more clearly, we have split into two the long
diagonal band along which all the models lie. The band on the left
shows the models with predominantly negative values of $\delta m^2$
(upper scale) which result in mostly negative values of $\Delta$ (left
hand scale). The band on the right includes the remaining models, {\it
i.e.} mainly models with positive values of $\delta m^2$ (lower scale)
which mostly yield $\Delta > 0$ (right hand scale). The horizontal bars
show the limits of the range of $\Delta$ within the mSUGRA model, 
$-5250 \ {\rm GeV}^2 \alt \Delta({\rm mSUGRA}) \alt -750 \ {\rm GeV}^2$.

Several features of this figure are worthy of note.
\begin{enumerate}
\item There is a very strong correlation between $\Delta$ and $\delta
m^2$. This is to be expected, since the first and last terms on the
right hand side of Eq.~(\ref{main}) are each proportional to $\delta m^2$
while the middle term is small and, as discussed above, approximately
constant. 

\item Eq.~(\ref{main}) suggests that the slope of the band would be
unity but for the last term that causes it to be smaller. It is easy to
check that for slepton masses $\sim 100$~GeV, the reduction in the slope
is about 8\% for each generation of sleptons with an intra-generational
splitting of $\delta m^2$. The slope of the band in the figure agrees
remarkably well with our expectation for three split slepton families. 

\item The width of the bands, {\it i.e.} the spread in the values of
$\Delta$, is essentially independent of $\delta m^2$. This is
understandable once we recognize that even in these models, the
replacement of $M_2$ by $m_{\tw_1}$ is the largest source of the spread
of $\Delta$. This is because the additional terms in Eq.~(\ref{main})
that were absent in mSUGRA are insensitive to model parameters. Clearly
the first term on the right hand side is completely independent, and the
last term is only logarithmically dependent on the slepton
mass.\footnote{To complete the argument we should also note that, since
slepton masses do not enter the chargino mass matrix, the modification
of allowing unequal slepton masses at the GUT scale does not
significantly alter $M_2-m_{\tw_1}$.}

\end{enumerate}

It is clear that if $\Delta$ lies significantly outside its
mSUGRA range, it should be possible to use it to distinguish models with
non-universal slepton masses from mSUGRA. The question then is, ``How
well can $\Delta$ be determined?''. Noting that the coefficient of the
$m_{\tw_1}^2$ term in the definition (\ref{delta}) of $\Delta$ is very
close to $1/2$, we can write the error in $\Delta$ as,
\begin{eqnarray*}
\delta\Delta &=&2\left[m_{\te_R}^2(\delta m_{\te_R})^2 +m_{\te_L}^2(\delta
m_{\te_L})^2 
+ {\frac{1}{4}}m_{\tw_1}^2(\delta m_{\tw_1})^2\right]^{{1}\over {2}} \\
&\simeq & 3\left({\frac{\delta m}{m}}\right) m^2,
\end{eqnarray*}
where in the last step we have assumed that the selectrons and charginos
have the same mass ($m$), and the same relative error in the mass
measurement. The {\it maximum} error in $\Delta$ which occurs when the
sparticles are all 250~GeV, is 1875~GeV$^2$, assuming that
selectron and chargino masses are measured with a precision of 1\%.  For
sparticle masses $\sim 150$~GeV, $\delta\Delta \simeq
675$~GeV$^2$. Allowing for a typical error $\sim 1000$~GeV$^2$ on
$\Delta$, we see from Fig.~\ref{nonuni}{\it a} that without any other
information it should be possible\footnote{To obtain this we linearly
combine the $1\sigma$ error in the measurement of $\Delta$ to the
uncertainty from the width of the band.} to detect intra-generational
non-universality of GUT scale slepton masses if $|\delta m^2| \agt
7000$~GeV$^2$.

It is, however, obvious that we can do better if we can reduce the
spread of $\Delta$ within the mSUGRA framework; this will also
correspondingly reduce the width of the band. A look at
Fig.~\ref{sugra}{\it b} and Fig.~\ref{sugra}{\it c} immediately shows
that the spread in $\Delta$ is considerably reduced once the chargino
mass is known. To illustrate this, keeping in mind that the chargino mass
can be determined \cite{okada,munroe,snow,zdr} at the percent level, we
plot $\Delta$ versus $\delta m^2$ in Fig.~\ref{nonuni}{\it b}, but only
for those models with $150 \leq m_{\tw_1} \leq 160$~GeV. The slanted
solid lines show the boundaries of the full band in frame {\it a}. We see
that using the chargino mass information\footnote{Reducing the mass
window further does not reduce the width of the band significantly, as
can be seen from Figs.~\ref{sugra}{\it b} and {\it c}.} together with
$\Delta$ should allow for detection of non-universality if $|\delta m^2|
\agt 5000$~GeV$^2$. For $m_0 \sim 120$~GeV, this corresponds to a
splitting of about 15\%. We have checked that unlike the chargino
mass, restricting the selectron mass to be close to its measured value
does not reduce the spread of the band, and so does not help to increase
the range over which non-universality might be detectable. Presumably,
this is because most of this spread comes from the difference $M_2 -
m_{\tw_1}$ which is insensitive to the selectron mass.

Can we improve upon this? Improving the precision with which sparticle
masses are measured will reduce the measurement error on $\Delta$ but
not the width of the band in Fig.~\ref{nonuni}{\it b}, and so does not
help a great deal as the latter is the major source of the uncertainty.
We see from the figure that the width is indeed reduced, along with the
mSUGRA range from Fig.~\ref{sugra}, if the sign of $\mu$
can be determined. There are mSUGRA studies \cite{okada,snow}
that suggest that this should be possible at linear colliders via
careful measurements of the chargino production channel. Although these
studies have only been done within the mSUGRA framework, the conclusions
regarding the properties of charginos and neutralinos that have been
obtained should also hold for the present case since the introduction of
$\delta m^2$ does not affect chargino and neutralino mass matrices
except via loop corrections. If the sign of $\mu$ is indeed determined,
the range of $\delta m^2$ over which the model with non-universality may
be confused with mSUGRA is reduced: For instance, for $\mu < 0$ and
$m_{\tw_1} \sim 150$~GeV, from Fig.~\ref{sugra}{\it c} and
Fig.~\ref{nonuni}{\it b}, we can infer that the two models should be
distinguishable except when $|\delta m^2| \alt
4000 \ {\rm GeV^2}$.

\begin{figure}
\dofig{6in}{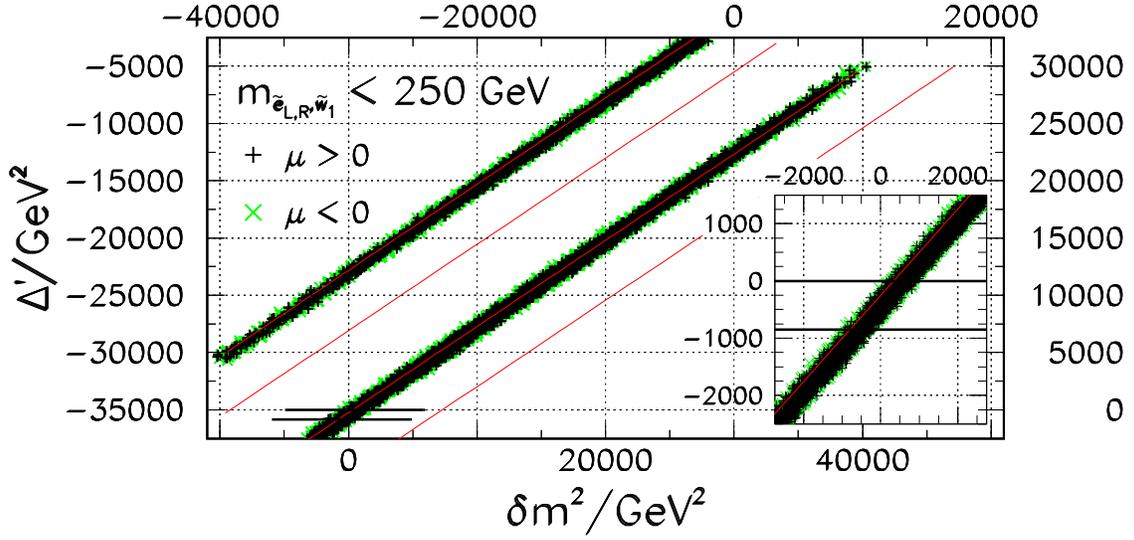}
\caption[]{A scatter plot of the quantity $\Delta'$ versus $\delta m^2$
defined by the left hand side of Eq.~(\ref{main}) for the same models as
in frame ({\it a}) of Fig.~\ref{nonuni}. The bands with postive and
negative values of $\mu$ essentially overlap so that the light crosses
are not visible. The slanted lines show the boundary of the bands in
Fig.~\ref{nonuni}{\it a}. The horizontal lines show the bounds on
$\Delta'$ within the mSUGRA framework. The inset shows a blow-up of the
neighborhood of the mSUGRA region. }
\label{m2}
\end{figure}

Yet another possibility for improving the sensitivity of $\Delta$ arises
if the gaugino mass parameter $M_2$ can directly be determined from the
measured chargino and neutralino masses. We would then not need the
replacement of $M_2$ by the chargino mass and the width of our band
would be greatly reduced. This is illustrated in Fig.~\ref{m2} where we
plot $\Delta'$, {\it defined} to be the left hand side of
Eq.~(\ref{main}), against $\delta m^2$ for the same models as in
Fig.~\ref{nonuni}{\it a}.  In this case, the positive and negative $\mu$
cases essentially overlap. The slanted lines are the boundaries of the
corresponding band in Fig.~\ref{nonuni}{\it a}. We see that the
$\Delta'$ band is shifted upwards relative to the $\Delta$ band. This is
because $M_2 \ge m_{\tw_1}$ for all these models.\footnote{This is not
true in general.}  We have checked that within the mSUGRA framework,
$\Delta'$ is negative and larger than $-850$~GeV$^2$, as shown by the
horizontal lines. While the greatly reduced mSUGRA range of $\Delta'$
makes it appear that we should be able to probe much smaller values of
$\delta m^2$, we should be cautious since measurement errors which were
previously smaller than the uncertainties from the width of the band may
now be the dominant limitation.

To assess the usefulness of $\Delta'$, we rely on previous studies that
examine how well the gaugino mass parameters may be determined. We
recognize that the extraction of the underlying gaugino masses could
depend on where we are in parameter space. Since only few case studies
have been performed, we should view the precision that we quote below
only as representative of what might be attainable. In their assessment
of how well gaugino mass unification could be tested at a linear
collider, Tsukamoto {\it et al.}  \cite{okada} have shown that even for
charginos as heavy as 220~GeV, it would be possible to measure $M_2$
with a precision of about 10\%, while the parameter $M_1$ could be
determined with a precision of 3\%. This was a result of a global fit to
the chargino and lightest neutralino masses, and slepton and chargino
production cross sections with polarized beams. Although they treated
$M_1$ and $M_2$ as independent parameters, assuming the gaugino mass
relation between these would imply that $M_2$ would also be determined
to at least 3\%. An mSUGRA case study done at the 1996 Snowmass
Workshop~\cite{snow} illustrates an example where $M_2$ is determined
with a precision of 1.5\%. Although this particular case which had a
chargino of just 97~GeV is now excluded by the LEP data, we present it
here to illustrate the precision that might be attainable at these
facilities. These studies all assume that just the lighter chargino is
kinematically accessible. If the heavier chargino is also accessible,
the analysis of Ref. \cite{choi} suggests that $M_2$ and  also $\mu$
might be extracted
with a precision of $\sim 1\%$, assuming only statistical errors
corresponding to an integrated luminosity of 1000~$fb^{-1}$. 
Precision measurements may also be possible at the
Large Hadron Collider, at least within the mSUGRA framework. Indeed
several cases are shown in Ref.\cite{ATLAS} where the unified gaugino
mass parameter is claimed to be determined with a precision of
1-2\%. For all the cases analyzed there, $m_{1/2}$ is claimed to be
determined to better than 10\%. In models with unified gaugino masses,
it is not unreasonable to suppose that $M_2$ would be determined with a
comparable precision. We stress though that these analyses determine
$m_{1/2}$ and not the gaugino masses independently, and also it is only
for favourable ranges of parameters that the very precise determination
of the gaugino masses is possible.

If slepton masses are determined to 1\%, and $M_2$ assumed to be
determined to (1,2,5,10)\%, then the corresponding error on $\Delta'$ is
about (700, 800, 1300, 2400) GeV$^2$ for $M_2$ and selectron masses of
150~GeV. We see from Fig.~\ref{m2} that a 1\% measurement of $M_2$
allows for a distinction between mSUGRA and the non-universal model for
$|\delta m^2|$ as small as 1900~GeV$^2$, while measurement of $M_2$ with
a precision of 5\% (10\%) degrades this to about 2700~GeV$^2$
(4100~GeV$^2$). It thus appears that an $M_2$ determination with a
precision of about 10\% is about as effective as using $\Delta$,
assuming that the sign of $\mu$ can be determined. It should, of course
be remembered that the uncertainty in $\Delta'$ scales quadratically as
the sparticle masses, so that the range of $\delta m^2$ over which
it will be possible to probe non-universality would be correspondingly
smaller (larger) if the sparticles are lighter (heavier) than 150~GeV.

We have attempted to leave our analysis as model independent as
possible. Nonetheless, it is interesting to ask what values $\Delta$ (or
$\Delta'$) might take in specific models, and whether these might be
probed by this type of analysis. We note that $\Delta$ in GMSB and AMSB
models differs qualitatively from its value within the mSUGRA
framework. But since these models have features that make them easily
distinguishable from mSUGRA, we focus on the ``much more mSUGRA-like''
models such as mSUGRA with scalar mass unification at $M_{Planck}$ or
gaugino-mediation. It is not our purpose to analyze these in detail
here, but we observe that in representative examples in the literature
(Fig.~3 of Ref.\cite{pp} or Fig.~1 of Ref.\cite{jhep}) a GUT scale
splitting $\sim 10-15$\% appears possible for SUGRA models with scalar
mass unification at $M_{Planck}$. For the gaugino-mediation model,
Fig.~6 of Ref.\cite{jhep} suggests that $\tell_L$ and $\tell_R$ masses
may be split by almost 20\% at $Q=M_{GUT}$.  Our analysis shows that
splittings of this magnitude should, quite possibly, be discernible in
experiments at the linear collider. 

Another model with split $\tell_L-\tell_R$ mass parameters that has been
recently examined is the $SO(10)$ SUSY GUT with $D$-terms. In this case,
$\delta m^2$ is an independent parameter equal to $4M_D^2$ in the
notation of Ref.\cite{soten}. In this study, it has been shown that
reasonable phenomenology can be obtained for $M_D \sim {{1}\over
{3}}m_0$. A slepton mass splitting of this magnitude should certainly be
detectable.

Before moving further, we mention that a recent study\cite{bm}
has claimed that by studying the threshold behaviour of the selectron
and chargino production cross sections, it should be possible to measure
these masses with a precision of $\sim 0.1$\%, assuming an integrated
luminosity of 100~$fb^{-1}$ for each measurement. The same study also
suggests that $M_2$ would be determined with a comparable
precision. Without making any representation about whether such
precision might actually be attainable, we merely note that if this
precision is
attained, it will be possible to probe slepton mass
non-universality for $|\delta m^2|$ as small as 1000~GeV$^2$ via a
determination of $\Delta'$. 

Up to now, our focus has been on whether models with non-universal
slepton masses can be distinguished from the mSUGRA model.  It should
be clear that experimental determination of $\Delta$ or $\Delta'$
can also directly be used to extract the slepton mass splitting parameter
$\delta m^2$. Just how well this can be done depends on what sparticle
masses turn out to be, since these reflect directly on the precision
with which $\Delta$ or $\Delta'$ can be determined. As an illustration,
we consider that the selectrons and the chargino each have a mass $\sim
150$~GeV, and that this is determined to within 1\%. In this case
$\Delta$ will be measured with an error of about $\pm 700$~GeV$^2$.
Fig.~\ref{nonuni}{\it b} then shows that the corresponding precision
with which $\delta m^2$ may be determined is about $\pm 2500$~GeV$^2$
(and somewhat better if the sign of $\mu$ is also determined).

The precision with which $\delta m^2$
might be determined via a measurement of $\Delta'$ is somewhat
better. For the same mass values as before, $\delta m^2$ may be
determined to within $\pm 1400$~GeV$^2$ ($\pm 2200$~GeV$^2$) if $M_2$
can be determined to within 1\% (5\%). Of course, since the $\Delta'$
band is much narrower, the actual precision will be more sensitive to
the actual values of sparticle masses which govern the error on
$\Delta'$.

We should compare our approach to (non-)universality to that in
Ref.\cite{zerwas}. In this study the values of sparticle masses
obtained from measurement were evolved to the GUT scale (using two loop
RGEs) to see if they converged to a common value within errors. This
approach requires that all sparticle masses be determined. For this, an
integrated luminosity of 1000~$fb^{-1}$ and a centre of mass energy up
to 1~TeV was required (to be able to measure squark masses which are
taken to be smaller than 500~GeV in this analysis). The main purpose
of this study was to examine how well the universality of scalar and
gaugino masses could be tested. It was assumed that first generation
slepton, squark and chargino masses would be measured by scanning the
threshold behaviour of the cross section with a precision of $\sim
0.1$\% \cite{bm}. It was shown that these measurements would
convincingly illustrate the underlying unification of scalar, as well as of
gaugino, masses at the GUT scale within the mSUGRA framework. It was not
the purpose there to examine how well these experiments could probe
non-universality and so, of course, there was no attempt to examine if
$\delta m^2$ could be determined.

Our study focuses on non-universality among the
sleptons, and makes no attempt to check the unification of slepton masses
with squark or Higgs boson masses. In contrast to Ref. \cite{zerwas},
measurements of just the two selectron and chargino masses are used in
our analysis. Finally, we also quantify how well the slepton mass splitting
at the GUT scale will be determined by these measurements.

\begin{figure}
\dofig{6in}{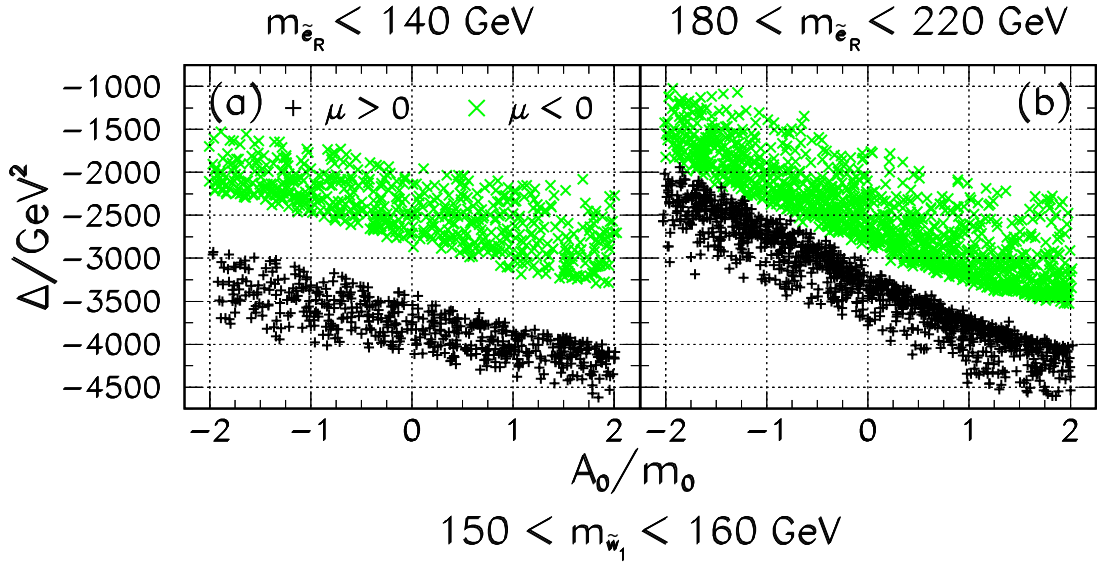}
\caption[]{A scatter plot of the quantity $\Delta$ versus $A_0/m_0$ for
mSUGRA models satisfying all constraints in Fig.~\ref{sugra}{\it b,c}
and for which $150 \ {\rm GeV} \leq m_{\tw_1} \leq 160$~GeV and ({\it
a})~100~GeV~$\leq m_{\te_R} \leq$~140~GeV, and ({\it b})~180~GeV~$\leq
m_{\te_R} \leq$~220~GeV. Models with a
positive value of $\mu$ are denoted by a dark plus while those with a
negative value of $\mu$ are denoted by a light cross.}
\label{A0}
\end{figure}

Before concluding, we point out one more feature about the quantity
$\Delta$ that is unrelated to the main subject of this paper. When
examining the correlation between $\Delta$ and other mSUGRA parameters,
we found that there is a surprising correlation with $A_0$. This is
illustrated in Fig.~\ref{A0} where we show $\Delta$ versus $A_0/m_0$ for
the set of mSUGRA models in Fig.~\ref{sugra}{\it b,c}, but with the chargino
mass in the range $150 \ {\rm GeV} \leq m_{\tw_1} \leq 160$~GeV, and
({\it a})~$m_{\te_R} \leq 140$~GeV, and ({\it b})~180~GeV $\leq
m_{\te_R} \leq$ 220~GeV. As before, models with positive (negative)
values of $\mu$ are denoted by a dark plus (light cross). Assuming
that sparticle masses are measured with a precision of 1\%, the
measurement error in $\Delta$ is expected to be $\sim 500$~GeV$^2$ and
$\sim 1200$~GeV$^2$, respectively. This is insufficient to pin down
$A_0/m_0$ with any precision. If, however, the results of the threshold
studies hold up, and mass measurements with an order of magnitude better
precision indeed turn out to be possible, then Fig.~\ref{A0}{\it b}
suggests that it may be possible to determine $A_0/m_0$ with a precision
of about $\pm {3\over 4}$ to $\pm 1$, if the selectron is heavy enough
and the sign of $\mu$ is known. While
this determination is crude, and only possible if sparticle masses happen
to lie in a favourable range, we thought it worthwhile to point it out
since we are not aware of any other way to get at this
parameter.\footnote{Measurements of third generation sparticle masses
may yield information about the corresponding parameter at the weak
scale, but the connection with $A_0$ is not clear.} 

We have checked that $\Delta'$ is relatively insensitive to
$A_0/m_0$. The correlation in Fig.~\ref{A0} arises because
$M_2-m_{\tw_1}$ is correlated to $A_0/m_0$. We have traced this to the
fact that the parameter $m_{H_u}^2$ in the renormalization group
improved 1-loop effective Higgs potential is strongly correlated with
$A_0/m_0$. The electroweak symmetry breaking condition then leads to a
correlation between $\mu$ (and since it appears in the chargino mass
matrix, also $m_{\tw_1}$) and $A_0/m_0$.  Indeed if $M_2$ and
$m_{\tw_1}$ can both be measured at the 0.1\% level, their mass
difference can be used to pin down $A_0/m_0$ with a precision
similar to that attained via a determination of $\Delta$ assuming $m_{\tw_1}$
and the two selectron  masses are measured to within 0.1\%. Of course, if
both $\tw_1$ and $\tw_2$ are accessible, then the extracted value of
$\mu$ \cite{choi} may provide a better handle on $A_0/m_0$.

{\it Summary}: We have studied whether experiments at a future
electron-positron collider operating at $\sqrt{s}=500$~GeV might be able
to probe an underlying non-universality in intra-generation slepton
masses at the GUT scale. Toward this end, we have identified a new
quantity $\Delta$ whose value can be directly determined from
experiments, and which, as can be seen in Fig.~\ref{nonuni} correlates
very strongly with $\delta m^2$, the difference of $\te_L$ and $\te_R$
mass squared parameters at the GUT scale. The value of $\Delta$ would be
very sensitive to the underlying mechanism of SUSY breaking, and would
differ dramatically between the mSUGRA, GMSB and possibly also AMSB
frameworks. Of course, these frameworks should be easy to distinguish
from one another once supersymmetric particles are discovered. The
point, however, is that $\Delta$ (or $\Delta'$) would be a
sensitive probe of slepton non-universality. We have shown that with a
1\% measurement of chargino and slepton masses, which is 
feasible with an integrated luminosity of
20-50~$fb^{-1}$, it should be possible to detect slepton mass
non-universality via a measurement of $\Delta$ if $|\delta m^2| \agt
5000$~GeV$^2$. If SUSY parameters are in a range that allow for a
determination of $M_2$ (or equivalently $M_1$) with a precision of 1-2\%
(5\%), then a determination of $\Delta'$ would allow non-universality to
be probed even if $|\delta m^2|$ is as small as 1900~GeV$^2$
(2700~GeV$^2$) for sparticle masses $\sim 150$~GeV. This should make it
possible to probe slepton mass non-universality as predicted by various
models in the literature. A measurement of the parameters $\Delta$ or
$\Delta'$ also allows a determination of $\delta m^2$ to a precision of
$\sim \pm 2500$~GeV$^2$, depending on model parameters.

\acknowledgments This research was supported in part by the
U.~S. Department of Energy under contract numbers DE-FG02-97ER41022 and
DE-FG-03-94ER40833. S.H. acknowledges financial support from the Deutsche
Forschungsgemeinschaft (DFG) under contract number HE 3241/1-1 that made
his stay in Hawaii possible.

%
%

\newpage
%
%



\end{document}